%% file: snrSAX.tex
\documentstyle{l-aa}
\input s_tables.tex

\newcommand{\be}{\begin{equation}}
\newcommand{\en}{\end{equation}}
\begin{document}
\def\ltsima{$\; \buildrel < \over \sim \;$}
\def\lsim{\lower.5ex\hbox{\ltsima}}
\def\gtsima{$\; \buildrel > \over \sim \;$}
\def\gsim{\lower.5ex\hbox{\gtsima}}
\def\spose#1{\hbox to 0pt{#1\hss}}
\def\approxlt{\mathrel{\spose{\lower 3pt\hbox{$\sim$}}
        \raise 2.0pt\hbox{$<$}}}
\def\approxgt{\mathrel{\spose{\lower 3pt\hbox{$\sim$}}
        \raise 2.0pt\hbox{$>$}}}
\def\deg {^\circ}
\def\mdot {\dot M}
\def\kms {~km~s$^{-1}$}
\def\gs {~g~s$^{-1}$}
\def\ergs {~erg~s$^{-1}$}
\def\cmtre {~cm$^{-3}$}\def\nupa{\vfill\eject\noindent}
\def\der#1#2{{d #1 \over d #2}}
\def\l#1{\lambda_{#1}}
\def\grb{$\gamma$-ray burst}
\def\grbs{$\gamma$-ray bursts}
\def\rosat{{\sl ROSAT} }
\def\cmdue {~cm$^{-2}$}
\def\gcm {~g~cm$^{-3}$}
\def\rsole{~R_{\odot}}
\def\msole{~M_{\odot}}
\def\aa #1 #2 {A\&A, {#1}, #2}
\def\mon #1 #2 {MNRAS, {#1}, #2}
\def\apj #1 #2 {ApJ, {#1}, #2}
\def\nat #1 #2 {Nature, {#1}, #2}
\def\pasj #1 #2 {PASJ, {#1}, #2}
\newfont{\mc}{cmcsc10 scaled\magstep2}
\newfont{\cmc}{cmcsc10 scaled\magstep1}
\newcommand{\bc}{\begin{center}}
\newcommand{\ec}{\end{center}}

\title{ Discovery of X--rays from the composite supernova remnant G0.9+0.1
with the $BeppoSAX$ satellite}
\author{ S. Mereghetti
\inst{1}, L. Sidoli\inst{1,2}, G.L. Israel\inst{3,4} 
}
\institute{
{Istituto di Fisica Cosmica del C.N.R., Via Bassini 15, I-20133 Milano,
Italy; \\ e-mail: (sandro, sidoli)@ifctr.mi.cnr.it}
\and
{Dipartimento di Fisica, Universit\`a di Milano, Via Celoria 16, I-20133
Milano, Italy}
\and
{Osservatorio Astronomico di Roma, Via dell'Osservatorio 2,
I-00040 Monteporzio Catone (Roma), Italy}
\and
{Affiliated to I.C.R.A.} \\
Accepted for publication on Astronomy and Astrophysics Letters}

\maketitle

\label{accepted sampout}

\begin{abstract}
 
Using the $BeppoSAX$ satellite we have obtained the first secure X--ray detection of the supernova remnant G0.9+0.1. The 1--10 keV spectrum can be described with an absorbed power law with $\alpha \sim$ 3 and $N_H\sim3\times10^{23}$\cmdue. The high column
 density supports a distance similar to that of the Galactic Center.
The X-ray emission, with a luminosity  
$L_{X}~\sim10^{35} d_{10}^{2} $ erg  s$^{-1}$,   
coincides with the
central radio core, confirming the composite nature of this remnant.
Though a search for periodic pulsations gave a negative result,
the observed X--rays are probably related to the presence of a 
young radio pulsar at the center  of G0.9+0.1.


\end{abstract}
 
\section{Introduction}
 
The radio source G0.9+0.1 was first recognized as a supernova remnant by 
Kesteven (1968) and later studied in detail with the VLA  and the 
Molonglo Observatory Synthesis Telescope. 
These observations clearly show that G0.9+0.1 consists
of a  steep--spectrum radio shell of $\sim8'$ diameter surrounding a core 
component with a flatter spectrum and significant polarization 
(Helfand \& Becker  1987; Gray 1994). 
Thus G0.9+0.1 belongs to the class of  "composite" supernova remnants,
that, in addition to the radio/X--ray shell formed by the expanding ejecta,
show the signature of a central neutron star powering   
non--thermal emission through the loss of rotational energy
(see Weiler  \& Sramek (1988) 
for a review of the classification of SNRs into shell--type,  plerionic, and composite).

Considering that the formation of a neutron star is expected
in most  supernovae (Branch 1990),   
it has been pointed out  that plerionic and composite
SNR should represent a larger fraction of the  known 
population of SNR than currently observed (Helfand et al.1989). 
It is not clear whether 
selection effects, that, especially in the radio band, favour the classification 
as SNR of shell sources, 
can entirely account for the observed discrepancy. 
X--ray observations are playing a crucial role in solving this problem,
both through the detection of new remnants
(Pfeffermann et al. 1991, Greiner et al. 1994,
Busser et al. 1996) and by showing that 
not all the neutron stars associated with SNR manifest themselves 
in the radio band as synchrotron nebulae and/or radio pulsars
(Mereghetti, Bignami \& Caraveo 1996;
Petre, Becker \& Winkler 1996; Vasisht \& Gotthelf 1997).
 
Until recently, most X-ray observations of SNR have been conducted below a few
keV, where interstellar absorption is severe and hampers the study of
distant objects in the galactic plane, such as G0.9+0.1. 
Though the distance of G0.9+0.1 is very uncertain, both the standard $\Sigma$--D analysis
(Downes 1971, Green 1991), and its  location near the galactic center direction,
indicate  a distance of $\sim$10 kpc or greater.
Despite the high absorption along this line of sight ($N_H\sim10^{23}$\cmdue),
the Italian--Dutch $BeppoSAX$ satellite, providing imaging capabilities
with arcminute angular resolution and  good sensitivity also above 4 keV, 
has  allowed a clear detection of the
X--ray emission from G0.9+0.1. 
 
\section{Observations and Data Analysis}
 
The   $BeppoSAX$ satellite (Boella et al. 1997a) carries a complement of
several imaging and non-imaging X--ray detectors, covering a broad
energy range from 0.1 keV to 300 keV.
Only the Medium and Low Energy Concentrator Spectrometers instruments  
are relevant for the results reported here. 
Both consist of position--sensitive gas--scintillation proportional counters, 
placed in the focal plane of the four coaligned  grazing incidence X--ray
mirrors carried by $BeppoSAX$.
The Medium-Energy Concentrator Spectrometer ($MECS$, Boella et al. 1997b) consists of three
identical  units covering the nominal energy range   1.3--10 keV. 
Its field of view has a diameter of $56'$. 
The Low Energy Concentrator Spectrometer ($LECS$, Parmar et al. 1997) has a thinner window
that allows to extend the low energy response to the 
range   0.1--10 keV. 
Both instruments have FWHM  energy resolution of  
$\sim8.5\sqrt{(6/{\rm E})}$\%, where E is the energy in keV. 
 
The location of G0.9+0.1 was imaged during a $BeppoSAX$ observation pointed on the SgrB2 
molecular cloud.
The observation was performed on April 5-6, 1997, and yielded net exposure times of 47 ks 
for each of the 3 $MECS$ units and   19 ks for the $LECS$.
Though G0.9+0.1 lies 14 arcmin off--axis, it was clearly detected in the $MECS$, 
with a total count rate of 0.023$\pm$0.002 counts s$^{-1}$.
The centroid of the X--ray emission ($R.A.=17h~47m~22s, 
Dec.=-28\deg~09'~28''$, J2000)
is consistent with the position of the  radio core of G0.9+0.1
(Helfand \& Becker 1987).

The on--axis angular resolution of the $MECS$ is $\sim2'$ (FWHM),
but it degrades to $\sim3'$ at the source off--axis location. 
Since the low statistics hampers a more detailed investigation of the source radial profile, 
to check whether the observed X--ray emission comes from
an extended source, we performed the following analysis. The ratio of the net counts within a 
radius  $R_{1}$ to those within a corona from $R_{1}$ to $R_{2}$ centered on the source 
position was compared to that expected from a point source. 
The reference data were obtained from ground based calibrations and from a  $\sim8$ ks exposure of Cyg X--1 
at the same    off--axis angle.
The comparison was done for different energy bands and different values of $R_{1}$ and $R_{2}$, 
always giving values compatible with emission from an unresolved source.


The $MECS$ counts used in the spectral analysis were extracted from circular regions  of $4'$ radius, and
the 256 original channels were grouped in order to have at least 20 net counts in each energy bin. 
The background spectrum was derived from a source free region of the same
observation. 

The spectrum is well fit by a simple power--law model, giving 
a reduced $\chi^2=0.95$ (42 d.o.f.) for photon index $\alpha=3.1$, $N_H=3.0\times 10^{23}$\cmdue, and 
flux $F = 2.12\times 10^{-11}$ erg cm$^{-2}$ s$^{-1}$  (1--10 keV, corrected for the absorption). 
Acceptable fits were also obtained with blackbody and thermal bremsstrahlung spectra
(see Table 1), while a Raymond-Smith thermal plasma model with abundances fixed at the solar values
gave a worse result.
There is no evidence for lines in the X-ray spectrum. In particular, we can set an upper limit 
of 400 eV (90\% confidence level) on the equivalent width of Fe K lines. 
 
The luminosity corresponding to the best fit power--law spectrum 
is $L_{X} = 2.4\times 10^{35} d_{10}^{2} $ erg  s$^{-1}$ in the 1--10 keV range,
but note that some uncertainty on this value is introduced by the relatively 
poorly constrained spectral parameters (see Table 1).
This is particularly evident for the power--law case 
([$0.8\times10^{35}, 1.6\times10^{36}$] $d_{10}^{2}$ erg  s$^{-1}$),
while the luminosity uncertainty is smaller for the other models.
 
G0.9+0.1 was also detected in the $LECS$ instrument, but only above $\sim$2 keV, 
due to  the high interstellar absorption. We verified that the inclusion of the $LECS$
counts in the spectral fits does not significantly change the
results described above.  
  

The same counts extracted for the spectral study were used for the timing analysis, 
after the correction of their times of arrival to the solar system barycenter. 
No flux variations were seen during the $BeppoSAX$ observation. 
A search for pulsations for periods in the range 8 ms to 2048 s gave a negative
result. With the hypothesis of a sinusoidal modulation, we can set the following upper limits on 
the pulsed fraction: 53\% for 8 $<$ P $<$ 16 ms, 38\% for 16 $<$ P $<$ 32 ms,
and 33\% for P $>$ 32 ms.

\begin{table*}
\label{spe}
\stablesthinline=0pt
\stablesborderthintrue
\stablestyle=0
\caption{{\bf} Results of the Spectral Fits (errors are 90\% c.l.).}
\begintable
Model          | Column density      | Parameter                           |Red. $\chi^2$   |Flux (1--10 keV)\el
               |($10^{22}$ cm$^{-2}$)|                                     |            |($10^{-11}$ ergs \cmdue s$^{-1}$ ) \el
 
Power law      |$30.3^{+14.7}_{-10.3}$        |$\alpha=3.1^{+1.4}_{-1.1}$           | 0.95 |     $2.1^{+12}_{-1.4}$      \el
Bremsstrahlung |$25.6^{+4.4}_{-5.6}$         |$T_{\rm br}=4.3^{+6.7}_{-1.3}$ keV    | 0.96 |     $0.94^{+0.40}_{-0.32}$   \el
Black body     |$17.9^{+7.1}_{-7.9}$          |$T_{\rm bb}=1.4^{+0.5}_{-0.3}$ keV   | 0.98 |     $0.42^{+0.24}_{-0.10}$   \el
Raymond-Smith  |$13.6^{+6.4}_{-3.6}$          |$T_{\rm RS}=31^{+\infty}_{-16}$ keV  | 1.22 |     $0.45$                 \endtable
\end{table*}

\section{Discussion}

Thanks to the  $BeppoSAX$ good sensitivity above a few keV,
our  observation has provided the first
firm evidence of  X--ray emission from G0.9+0.1.
Though this region of sky has been imaged in the past with other X--ray satellites,
only a very marginal and uncertain detection of G0.9+0.1
was reported by Helfand \& Becker (1987),
based on an observation done  with the 
Imaging Proportional Counter (IPC) on the Einstein Observatory in 1979. 
The claimed detection  of a source with  0.009$\pm$0.003 counts s$^{-1}$
at more than one arcminute west of the SNR center,
was based on a rather {\it ad hoc} procedure aimed at maximizing 
the very small number of net counts and on an uncertain background estimate.
Reanalysing the same IPC pointing (5 ks),  we found 
no sources above a signal to noise ratio of 2, inside the region
corresponding to the  radio SNR. 
With these data we can only put an upper limit of  0.03 counts s$^{-1}$,
while the expected IPC count rate, based on our best fits, is only of $\sim10^{-3}$ counts s$^{-1}$,
one order of magnitude below that reported by Helfand \& Becker (1987).


A search in the ROSAT public archives yielded several PSPC and HRI observations 
containing the position of G0.9+0.1.
However, due to the short exposure times of
only a few thousand seconds and especially to the high absorption in the ROSAT band, the SNR was not detected.
 
All our spectral fits give values of $N_H$ greater than $10^{23}$\cmdue,
indicating that G0.9+0.1 must be at a distance of several kiloparsecs,  
probably close to the Galactic Center or even beyond it. In the following
discussion we shall assume a distance of 10 kpc. 
The large interstellar absorption also explains the apparent discrepancy between
our derived luminosity of a few $10^{35}$ erg  s$^{-1}$
and the smaller one estimated by Helfand \& Becker (1987),
who assumed a lower $N_H$ value for the IPC count rate to flux conversion.

The peak of the X--ray emission is coincident with the SNR radio core  
and there is no evidence for a spatial extension greater than the instrumental
resolution.
Were the X--rays emitted from a shell with the same dimensions observed in the 
radio band (diameter $\sim8$ arcmin), they would appear clearly resolved
in the $MECS$ images. 
Therefore, we  are clearly seeing X--rays emitted 
predominantly from the central region of the remnant,
either from a point source or from a nebula with radius smaller than $\sim2$ arcmin.
  
Some SNRs, like for example W44 (Rho et al. 1994), present a centrally peaked X--ray emission 
of thermal origin. The thermal nature of the emission
is clearly demonstrated by the detection of lines in
their X--ray spectra. 
All the SNRs of this kind have a limb-brightened radio
morphology  without  a flat--spectrum core, contrary to the case of G0.9+0.1. 
Also considering that  the thermal plasma model
gave the worst fit to our  data,   
we favour the alternative interpretations
related to the likely presence of a neutron star
at the center of G0.9+0.1.

One possibility is that of thermal emission from the neutron star
surface. The results of the blackbody spectral fit imply an emitting surface with radius
$R=0.3^{+0.4}_{-0.2}  d_{10}$ km,  
definitely smaller than the whole neutron star surface for any reasonable distance.
This can be interpreted as emission from a small polar cap
region, hotter than the rest of the neutron star due to anisotropic
heat diffusion from the interior and/or to reheating by relativistic particles  
backward accelerated in the magnetosphere
(Halpern \& Ruderman 1993).
In general, this should produce a periodic flux modulation, but 
the  strong gravitational bending
effects severely reduce the observed pulsed fractions 
(Page 1995).
Our upper limits on the possible flux modulations are not 
strong enough to pose serious problems to this interpretation.
However, the fitted blackbody temperature (kT$\sim$1.4 keV) is higher than that
observed in all the other X--ray emitting radio pulsars.

A different explanation  involves non-thermal emission powered by the rotational energy loss
of a relatively young  neutron star.
The radio shell radius of $\sim$12 pc implies a lower limit to the remnant age of $\sim$1100 yr,
for a free-expansion phase with   v$\sim10^{4}$\kms.  
If the remnant is expanding adiabatically, from the Sedov model we have a shell radius 
$R\sim14(E_{51}/n_{o})^{1/5}t_{4}^{2/5}$pc, where $E_{51}$ is the explosion energy 
in units of $10^{51}$ ergs, $n_{o}$ is the ambient ISM hydrogen density in $cm^{-3}$
and  $t_{4}$ is the age in units of $10^{4}$ yr. 
For typical values $(E_{51}/n_{o})^{1/5}\sim 1$, we derive an age  of $\sim 6,800$ years. 
Both a point--like, pulsed component originating in the neutron star magnetosphere
and a diffuse ($1-3'$) synchrotron nebula probably contribute to the observed X--rays.

Our best fit power law photon index 3.1 is rather steep, compared to  
other X--ray synchrotron  nebulae, 
but a more typical value of $\alpha=2$ is also consistent with our data  (for   
$N_H=2\times 10^{23}$\cmdue). 
The corresponding X--ray luminosity (1-10 keV), $L_{X}=8.2\times 10^{34} d_{10}^{2} $ erg s$^{-1}$, 
is within the range observed in the central components of other SNRs
(see, e.g., Helfand \& Becker 1987) and can be easily powered by a young
neutron star.

 \section{Conclusions}

With the $BeppoSAX$ satellite we have discovered X--ray emission from the central region of the supernova remnant G0.9+0.1
located close to the Galactic Center direction. 
The high interstellar absorption is consistent with a distance of the order of 10 kpc
and, correspondingly, an X--ray luminosity  of  $\gsim10^{35}$ erg  s$^{-1}$.
 
Although we cannot completely rule out a thermal origin of the X--ray emission,
its small angular extent (radius $\lsim2'$), the good fit with a power-law,
the presence of a flat spectrum radio core, and the estimated SNR age of
a few thousand years, favour the interpretation
in terms of  synchrotron emission powered by a young, energetic pulsar.

\begin{acknowledgements}
We thank Lucio Chiappetti and Silvano Molendi for useful discussions and help with the data analysis.

\end{acknowledgements}

\end{document}

%% file: s_tables.tex
\newhelp\stablestylehelp{You must choose a style between 0 and 3.}%
\newhelp\stablelinehelp{You should not use special hrules when stretching
a table.}%
\newhelp\stablesmultiplehelp{You have tried to place an S-Table inside another
S-Table.  I would recommend not going on.}%
\newdimen\stablesthinline
\newdimen\stablesthickline
\newif\ifstablesborderthin
\stablesborderthinfalse
\newif\ifstablesinternalthin
\stablesinternalthintrue
\newif\ifstablesomit
\newif\ifstablemode
\newif\ifstablesright
\stablesrightfalse
\newdimen\stablesbaselineskip
\newdimen\stableslineskip
\newdimen\stableslineskiplimit
\newcount\stablesmode
\newcount\stableslines
\newcount\stablestemp
\newcount\stablescount
\newcount\stableslinet
\newcount\stablestyle
\def\stablesleft{\quad\hfil}%
\def\stablesright{\hfil\quad}%
\catcode`\|=\active%
\newcount\stablestrutsize
\newbox\stablestrutbox
\setbox\stablestrutbox=\hbox{\vrule height10pt depth5pt width0pt}
\def\stablestrut{\relax\ifmmode%
                         \copy\stablestrutbox%
                       \else%
                         \unhcopy\stablestrutbox%
                       \fi}%
\newdimen\stablesborderwidth
\newdimen\stablesinternalwidth
\newdimen\stablesdummy
\newcount\stablesdummyc
\newif\ifstablesin
\stablesinfalse
\def\begintable{\stablestart%
  \stablemodetrue%
  \stablesadj%
  \halign%
  \stablesdef}%
\def\stablesadj{%
  \ifcase\stablestyle%
    \hbox to \hsize\bgroup\hss\vbox\bgroup%
  \or%
    \hbox to \hsize\bgroup\vbox\bgroup%
  \or%
    \hbox to \hsize\bgroup\hss\vbox\bgroup%
  \or%
    \hbox\bgroup\vbox\bgroup%
  \else%
    \errhelp=\stablestylehelp%
    \errmessage{Invalid style selected, using default}%
    \hbox to \hsize\bgroup\hss\vbox\bgroup%
  \fi}%
\def\stablesend{\egroup%
  \ifcase\stablestyle%
    \hss\egroup%
  \or%
    \hss\egroup%
  \or%
    \egroup%
  \or%
    \egroup%
  \else%
    \hss\egroup%
  \fi}%
\def\stablestart{%
  \ifstablesin%
    \errhelp=\stablesmultiplehelp%
    \errmessage{An S-Table cannot be placed within an S-Table!}%
  \fi
  \global\stablesintrue%
  \global\advance\stablescount by 1%
  \message{<S-Tables Generating Table \number\stablescount}%
  \begingroup%
  \stablestrutsize=\ht\stablestrutbox%
  \advance\stablestrutsize by \dp\stablestrutbox%
  \ifstablesborderthin%
    \stablesborderwidth=\stablesthinline%
  \else%
    \stablesborderwidth=\stablesthickline%
  \fi%
  \ifstablesinternalthin%
    \stablesinternalwidth=\stablesthinline%
  \else%
    \stablesinternalwidth=\stablesthickline%
  \fi%
  \tabskip=0pt%
  \stablesbaselineskip=\baselineskip%
  \stableslineskip=\lineskip%
  \stableslineskiplimit=\lineskiplimit%
  \offinterlineskip%
  \def\borderrule{\vrule width \stablesborderwidth}%
  \def\internalrule{\vrule width \stablesinternalwidth}%
  \def\thinline{\noalign{\hrule height \stablesthinline}}%
  \def\thickline{\noalign{\hrule height \stablesthickline}}%
  \def\trule{\omit\leaders\hrule height \stablesthinline\hfill}%
  \def\ttrule{\omit\leaders\hrule height \stablesthickline\hfill}%
  \def\tttrule##1{\omit\leaders\hrule height ##1\hfill}%
  \def\stablesel{&\omit\global\stablesmode=0%
    \global\advance\stableslines by 1\borderrule\hfil\cr}%
  \def\el{\stablesel&}%
  \def\elt{\stablesel\thinline&}%
  \def\eltt{\stablesel\thickline&}%
  \def\elttt##1{\stablesel\noalign{\hrule height ##1}&}%
  \def\elspec{&\omit\hfil\borderrule\cr\omit\borderrule&%
              \ifstablemode%
              \else%
                \errhelp=\stablelinehelp%
                \errmessage{Special ruling will not display properly}%
              \fi}%
  \def\stmultispan##1{\mscount=##1 \loop\ifnum\mscount>3 \stspan\repeat}%
  \def\stspan{\span\omit \advance\mscount by -1}%
  \def\multicolumn##1{\omit\multiply\stablestemp by ##1%
     \stmultispan{\stablestemp}%
     \advance\stablesmode by ##1%
     \advance\stablesmode by -1%
     \stablestemp=3}%
  \def\multirow##1{\stablesdummyc=##1\parindent=0pt\setbox0\hbox\bgroup%
    \aftergroup\emultirow\let\temp=}
  \def\emultirow{\setbox1\vbox to\stablesdummyc\stablestrutsize%
    {\hsize\wd0\vfil\box0\vfil}%
    \ht1=\ht\stablestrutbox%
    \dp1=\dp\stablestrutbox%
    \box1}%
  \def\stpar##1{\vtop\bgroup\hsize ##1%
     \baselineskip=\stablesbaselineskip%
     \lineskip=\stableslineskip%
     \lineskiplimit=\stableslineskiplimit\bgroup\aftergroup\estpar\let\temp=}%
  \def\estpar{\vskip 6pt\egroup}%
  \def\stparrow##1##2{\stablesdummy=##2%
     \setbox0=\vtop to ##1\stablestrutsize\bgroup%
     \hsize\stablesdummy%
     \baselineskip=\stablesbaselineskip%
     \lineskip=\stableslineskip%
     \lineskiplimit=\stableslineskiplimit%
     \bgroup\vfil\aftergroup\estparrow%
     \let\temp=}%
  \def\estparrow{\vfil\egroup%
     \ht0=\ht\stablestrutbox%
     \dp0=\dp\stablestrutbox%
     \wd0=\stablesdummy%
     \box0}%
  \def|{\global\advance\stablesmode by 1&&&}%
  \def\|{\global\advance\stablesmode by 1&\omit\vrule width 0pt%
         \hfil&&}%
  \def\vt{\global\advance\stablesmode by 1&\omit\vrule width \stablesthinline%
          \hfil&&}%
  \def\vtt{\global\advance\stablesmode by 1&\omit\vrule width \stablesthickline%
          \hfil&&}%
  \def\vttt##1{\global\advance\stablesmode by 1&\omit\vrule width ##1%
          \hfil&&}%
  \def\vtr{\global\advance\stablesmode by 1&\omit\hfil\vrule width%
           \stablesthinline&&}%
  \def\vttr{\global\advance\stablesmode by 1&\omit\hfil\vrule width%
            \stablesthickline&&}%
  \def\vtttr##1{\global\advance\stablesmode by 1&\omit\hfil\vrule width ##1&&}%
  \stableslines=0%
  \stablesomitfalse}
\def\stablesdef{\bgroup\stablestrut\borderrule##\tabskip=0pt plus 1fil%
  &\stablesleft##\stablesright%
  &##\ifstablesright\hfill\fi\internalrule\ifstablesright\else\hfill\fi%
  \tabskip 0pt&&##\hfil\tabskip=0pt plus 1fil%
  &\stablesleft##\stablesright%
  &##\ifstablesright\hfill\fi\internalrule\ifstablesright\else\hfill\fi%
  \tabskip=0pt\cr%
  \ifstablesborderthin%
    \thinline%
  \else%
    \thickline%
  \fi&%
}%
\def\endtable{\advance\stableslines by 1\advance\stablesmode by 1%
   \message{- Rows: \number\stableslines, Columns:  \number\stablesmode>}%
   \stablesel%
   \ifstablesborderthin%
     \thinline%
   \else%
     \thickline%
   \fi%
   \egroup\stablesend%
\endgroup%
\global\stablesinfalse}